\documentstyle[aps,preprint]{revtex}
\begin{document}
\draft
\tightenlines

\title{Nonequilibrium Phase Transitions in Epidemics and Sandpiles}

\author{Ronald Dickman$^{*}$}
\address{
Departamento de F\'{\i}sica, ICEx,
Universidade Federal de Minas Gerais,\\
30123-970
Belo Horizonte - MG, Brasil\\
}
\date{\today}

\maketitle
\begin{abstract}
Nonequilibrium phase transitions between an active and an absorbing state  
are found in models of populations,
epidemics, autocatalysis, and chemical reactions on a surface. 
While absorbing-state phase transitions fall generically in the DP 
universality class, this does not preclude other universality 
classes, associated with a symmetry or 
conservation law. An interesting issue concerns the dynamic 
critical behavior of models with an infinite number of absorbing 
configurations or a long memory.  Sandpile models, the principal
example of self-organized criticality (SOC), also exhibit absorbing-
state phase transitions, with SOC corresponding to a particular mode 
of forcing the system toward its critical point.  
\vspace{1em}

\end{abstract}

\section{Introduction}

Phase transitions arise not only in thermodynamic equilibrium, but 
in nonequilibrium situations such as the onset of convection \cite{ahlers},
and in models of populations, epidemics, catalysis, cooperative transport, 
and markets, to cite but a few examples. 
The simplest models of spatially extended, nonequilibrium systems are 
lattice Markov processes,
or interacting particle systems\cite{liggett}.
There is usually no Hamiltonian; rather, the model is defined by a set of transition 
rates, that do not satisfy detailed balance with respect to a
`reasonable' energy function.

Despite these differences from equilibrium statistical mechanics,
phase transitions (which mark a singular dependence of macroscopic 
properties on control parameters), scaling, and universality, retain their
fundamental importance.  Such concepts
transcend the equilibrium/nonequilibrium boundary.
In fact, familiar universality classes, such Ising or Potts, still appear
in the nonequilibrium context, along with others not encountered
in equilibrium.
While new classes are still being identified, and the general
principles determing critical behavior far from equilibrium have yet to be
specified completely, it has become clear that symmetry and conservation
laws play a fundamental role, just as in equilibrium.
Some useful reviews of various aspects of nonequilibrium phase transitions
can be found in Refs. \cite{schmitt,marro,hinr,schutz}.  Here I discuss
some recent work on absorbing-state phase transitions and their relation to
self-organized criticality.

\section{Absorbing-state phase transitions}

Absorbing-state phase transitions arise from a conflict between the spread of
activity, and a tendency for this activity to
die out \cite{marro,hinr}.  The activity may represent an epidemic \cite{harris}, a reaction
proceeding on the surface of a catalyst \cite{zgb}, or spatiotemporal
chaos \cite{pomeau,bohr}, for example.
When the system is translation-invariant, with interactions of finite range,
and there are no conserved quantities, the critical behavior is
expected to fall generically in the directed-percolation (DP) universality class
\cite{janssen,pg82}, as has been verified in studies of a wide variety of systems.

DP and the closely related contact process \cite{harris} exhibit well known
scaling properties, for example the stationary order parameter (activity density)
grows $\propto (\lambda - \lambda_c)^{\beta}$, where $\lambda$ is the
spreading or reproduction rate for activity.  At the critical point, $\lambda_c$, starting from
a localized active region (e.g., a single active site), the survival probability $P(t)$
decays $ \propto t^{-\delta} $, while the mean number of active sites $n(t)$
grows $\propto t^{\eta}$; these scaling laws describe avalanches of 
activity \cite{avexp}.  For the DP class, a field theory, framed in terms of a
nonnegative, scalar order parameter density $\rho(x,t)$ yields an upper critical
dimension of four, and an $\epsilon$-expansion for the critical exponents
\cite{janssen,pg82,cardy}:

\begin{equation}
\frac{ \partial \rho}{\partial t} = \nabla^2 \rho -a \rho - b\rho^2 + \eta(x,t) \;.
\label{cpft}
\end{equation}

\noindent Here $\eta (x,t)$ is zero-mean Gaussian noise 
with autocorrelation

\begin{equation}
\langle \eta(x,t) \eta(x',t') \rangle = \Gamma \rho(x,t) \delta(x-x') \delta(t-t') \;.
\label{noisecp}
\end{equation}

Some other universality classes associated with absorbing-state phase transitions
are: compact DP, in which activity is bounded by random walks
\cite{dk,essam}; the parity-conserving or directed-Ising class, exemplified by
branching-annihilating random walks with an even number of offspring
\cite{gkt,tt,cardy96}; and the ``conserved nondiffusing field" class
\cite{rossi}.  The criteria for a model to belong to the parity-conserving
universality class are still not completely clear.

\subsection{Interface representation}

A recent development in this area is the application of surface-growth scaling ideas.
Consider the contact process, in
which, at a given instant, each site of the lattice is either occupied ($\sigma_i = 1$),
or vacant ($\sigma_i = 0$).  Transitions from $\sigma_i = 1$ to $\sigma_i = 0$
occur at a rate of unity, independent of the neighboring sites.  The
reverse can only occur if at least one neighbor is occupied: the transition
from $\sigma_i = 0$ to $\sigma_i = 1$ occurs at a rate of $\lambda m$,
where $m$ is the fraction of nearest neighbors of site $i$ that are occupied;
thus the state $\sigma_i = 0$ for all $i$ is absorbing.  The stationary 
order parameter (the activity density, or fraction of occupied sites), is zero 
for $\lambda < \lambda_c$.

We define a set of height variables

\begin{equation}
h_i (t) = \int_0^t \rho_i (t') dt' 
\end{equation}
that represent the cumulative activity at each site.  Since the $h_i$ are
nondecreasing functions of time, we can think of them as describing a
a growing or driven interface, and introduce the width $W$
via $W^2 (t,L) = \mbox{var}[h_i (t)]$, on a lattice of $L^d$ sites.
The width exhibits Family-Vicsek scaling \cite{fv,barabasi},
but with the important difference that here it saturates only
when the process falls into the absorbing state.

At the critical point,
the scaling exponents for the width are related to those of DP.  For example,
the growth exponent (defined via $W \sim t^{\beta_W}$)
satisfies $\beta_W + \delta = 1$,
where $\delta$ describes the decay of activity at the critical point,
starting with all sites active: $\rho \sim t^{-\delta}$.  This relation
follows from a scaling hypothesis for the 
probability distribution $P(h;t)$ of height $h$ (at any site): 

\begin{equation}
P(h;t) = \frac{1}{\overline{h}(t)} {\cal P} [h/\overline{h}(t)] ,
\end{equation}
(${\cal P} $ is a scaling function, $\overline{h}(t)$ the mean height), 
that has been verified in simulations 
\cite{intcp}.  Studies of the Domany-Kinzel model show that the
growth exponent exhibits a sharp cusp at the critical point, making it
a sensitive indicator of criticality \cite{allbens}. 

\section{How sandpiles work}

{\it Self-organized criticality} (SOC) is usually taken to mean ``spontaneous"
criticality, or scale-invariance in the absence of identifiable tuning of 
parameters \cite{btw}.
In sandpiles, SOC is an absorbing-state phase transition of the same kind
as in DP (though with different critical exponents), subject to an external
control mechanism \cite{dvz,socbjp}.  A simple example illustrates how this
operates.

Consider a system of {\it activated random walkers} (ARW) on a lattice with periodic
boundaries in $d \geq 1$ dimensions.  $N$ walkers are initially distributed 
at random over the $L^d$ sites, with no restriction on the number of walkers per site.
The dynamics, which conserves the number of walkers, is simply that each
site with two or more walkers has a rate of 1 to liberate a pair of walkers.
Each liberated walker moves independently, without bias, to one of the neighboring sites 
(in one dimension, from site $j$ to $j+1$ or $j-1$ with equal likelihood).
Isolated walkers are however paralyzed.
The system has two kinds of configurations: active, in which at least
one site has $\geq 2$ walkers, and absorbing, with all walkers immobile.  
For $N > L^d$ only active configurations are possible, and activity continues forever.  

For $N \leq L^d$ there are both active and
absorbing configurations, the latter representing a shrinking fraction of 
configuration space as the density $\zeta \equiv N/L^d \rightarrow 1$.  Given
an active initial configuration (a virtual certainty for a random distribution with $\zeta 
>0$ and $L$ large), will the system remain active indefinitely, or will it fall into an absorbing 
configuration?  For small $\zeta$ it should be easy for the latter to occur, but it seems 
reasonable that for sufficiently large densities (but $ < 1$),
the likelihood of reaching an absorbing configuration becomes so 
small that the system remains active indefinitely.
Mean-field theory  and extensive simulations confirm the existence of
sustained activity for densities greater than
some critical value $\zeta_c$, with $\zeta_c < 1$
\cite{socbjp,fes2d,manna1d}. 
The former predicts $\zeta_c=1/2$ while
simulation yields 0.9489 for the critical density in one dimension,
0.7170 in two dimensions \cite{rnote}.
Thus the ARW model exhibits a continuous absorbing-state phase transition
at a critical density $\zeta_c$.  The critical exponents are {\it not} those of DP,
due to a conserved density (i.e., of walkers), which relaxes diffusively
in the presence of activity, but is frozen in inactive regions \cite{rossi,granada}.

Now we make two simple changes in the model:
\vspace{1em}

i) We open the boundaries; walkers can now jump out from the edge.
\vspace{1em}

ii) To compensate for the loss of walkers at the boundaries, we include a
source, which inserts a new walker at a random site,
if and only if the system has fallen into an absorbing configuration.
(Note that the source enjoys {\it global} information regarding the state of the system.)
\vspace{1em}

It is easy to see that these innovations amount to a control mechanism that
forces the system to its critical point.  Suppose that $\zeta$ is above the critical value;
then there is activity (hence no addition), and loss of walkers when activity reaches the
boundary, so that $d \zeta /dt < 0$.  If $\zeta < \zeta_c$, on the other hand, there is
no activity (and no loss), and the source is actuated, making $d \zeta /dt > 0$. 
Thus the system is forced to the critical point, and, quite naturally, exhibits scale
invariance in the stationary state.  Thus modified, the activated random walkers
model is a continuous-time version of Manna's stochastic sandpile \cite{manna};
the loss and addition mechanisms yield what is commonly called self-organized
criticality.  Their functioning depends on the conservation of walkers, allowing the
control mechanism to be described in terms of insertion and loss of ``sand",
 without ever making reference to the underlying control parameter $\zeta$.

Studies of criticality in sandpiles with a periodic boundaries (so-called fixed-energy
sandpiles), have yielded their critical exponents and associated avalanche exponents \cite{fes2d,manna1d}, and have suggested a field theory for sandpiles \cite{sandft}.
The theory involves two fields, the activity density $\rho_a$ and the particle
density $\zeta$.  Upon formal elimination of the latter, one obtains

\begin{eqnarray}
\frac{\partial \rho_a ( {\bf x},t)}{\partial t}
& =& D_a\nabla^2 \rho_a({\bf x},t) - r ({\bf x}) \rho_a({\bf x},t)
 - b \rho_a^2({\bf x},t)   \nonumber \\
& + & w \rho_a({\bf x},t) \int_0^t dt' \nabla^2 \rho_a({\bf x},t')
  + \sqrt{\rho_a} \eta ( {\bf x},t).
\label{main}
\end{eqnarray}
$\eta$ is a Gaussian white noise with autocorrelation $\langle \eta(  {\bf x},t) \eta ( {\bf x'},t')\rangle = D \delta({\bf x-x'})
\delta(t-t')$; $c, b$ and $w$ are fixed parameters.  The theory resembles
that for DP, except for the Laplacian memory term, and the spatial
dependence of the growth rate $r$, inherited from the initial configuration.
This theory has yet
to yield to renormalization group analysis \cite{granada}.

Simulations support the proposal  \cite{narayan,pacz,alava}
that stochastic sandpiles fall in the same universality class as that of depinning of a 
linear elastic interface subject to random pinning forces, at least for $d \geq 2$.  The
deterministic Bak-Tang-Wiesenfeld sandpile \cite{btw}, by contrast, appears to define a universality class {\it sui generis}, and is marked by strong nonergodic effects.

We close this section with some observations on SOC itself.  SOC was proposed as
a paradigm for explaining the appearance of scale invariance in nature without the need
to invoke an outside agent tuning the system to criticality \cite{btw,hnw}.  We have seen, however, that the source [component ii) of the control scheme], will require a huge quantity of
information ($L^d$ bits) in order to decide whether or not to add sand, so that the riddle
of scale invariance in nature seems not to have been resolved.  A way out of this
dilemma is to suppose that the source is does not have access to all this
information, but acts on a time scale much longer than the local
dynamics.  Then we can expect to observe scale invariance over a considerable range
of avalanche sizes and durations, large enough, in principle, to agree with observation. 
A similar consideration applies to strict conservation: admitting a small dissipation rate,
certain sandpile models exhibit ``quasi-criticality" which may again be sufficient to
account for observed scale invariance \cite{prado}.

\section{Infinite numbers of absrobing configurations}

Infinite numbers of absorbing configurations
arise in models of catalysis \cite{dt}, and, as noted above,
in sandpiles.  The simplest model with an infinite number of absorbing configurations
is Jensen's  pair contact process (PCP) \cite{pcp1,pcp2}.
In this model, a pair of particles annihilates with probability $p$, and creates
a new particle (at a vacant neighbor site) with probability $1\!-\!p$.
Any configuration devoid of nearest-neighbor pairs is absorbing.

The static critical behavior of the PCP falls in the directed percolation universality
class, but in one dimension the spreading exponents $\delta$ and $\eta$ 
{\it vary continuously} with the particle
density $\phi$ in the environment \cite{pcp2,pcp3}.
The spreading exponents take their DP values (e.g., $\delta = 0.16$),
only when the initial density $\phi$ of isolated
particles is set to the ``natural" value $\phi_{nat} \simeq 0.242$,
i.e., the value generated by the process itself, running at the critical point.
(For $\phi = 0$, for example, $\delta \simeq 0.27$.)
Numerically, $p_c$,
$\delta + \eta$, and $z$ (the exponent governing the spread of the
active region: $R^2 \sim t^z$), appear to be independent of $\phi$.
Exponents $\delta$, $\eta$, and $z$ satisfy a generalized hyperscaling relation
\cite{mendes,yu}.

The issue of variable spreading exponents remains unresolved and controversial.
While evidence for variable exponents has been found in many simulations
\cite{bohr,pcp2,pcp3,odor98,lipowski,odor00}, there is as yet no 
complete theory.  A field theory
\cite{inas} again involves the order parameter $\rho$ coupled to a second
field (which may represent the density of isolated particles),
that is frozen in regions with $\rho = 0$:

\begin{equation}
\frac{\partial \rho}{\partial t} =
D_ \rho \nabla^2 \rho - a_\rho \rho 
 -b_\rho   \rho^2     
+ w_\rho  \phi \rho
+ \eta_\rho
\end{equation}
 
\begin{equation}
\frac{\partial \phi}{\partial t} = D_\phi  \nabla^2  \rho
 - a_\phi  \rho - b_\phi  \rho^2
+w_\phi  \phi \rho 
+ \eta_\phi
\end{equation}
(Here both noise autocorrelations are $\propto \rho$.)
Eliminating $\phi$, one obtains an equation for $\rho$ with a memory term;
the stationary properties have been shown to be the same as DP.  While it has not
been possible to determinine the spreading exponents analytically,
numerical integration of the field theory
yields $\delta$ and $\eta$ that vary with the coupling $w_\rho$ 
between the two fields \cite{lopez}.

The difficulty of understanding nonuniversal spreading
in models with an infinite number of absorbing configurations
has motivated the study of simplified descriptions.
Noting that anomalous spreading behavior is associated with a
long memory of the initial configuration,
Grassberger et al. proposed a {\it generalized epidemic process},
in which DP (with a unique absorbing configuration), 
is modified to have long memory \cite{gcr}.  Specifically, the
probability $p'$ for a virgin site to become infected is different
from that ($p$) for used sites (those that have already been infected
and have recovered).
These authors find  that in one dimension, at the critical point $p_c$
(which is independent of $p'$), the  model exhibits variable 
exponents $\delta$ and $\eta$ for $p' > p$, but
{\it faster than power-law decay} for $p' <p$, that is, no scaling. 

Since the very existence of scaling in a model with a long memory of the initial
condition appears to be in doubt, it is helpful to have some simple
examples in which a variable exponent can be established analytically.
A continuously variable survival probability exponent, $\delta$, has been
demonstrated for DP confined to a parabola \cite{turban}, and for
compact DP in a similar geometry \cite{odor}.  
In these cases the space-time boundary is fixed, but
recently a similar result was shown 
for a random walk with a long memory in the form of a
movable partial reflector \cite{rwmpr}.

Consider an unbiased, discrete-time random walk $x_t$ on a one-dimensional 
lattice, with $x_0$ = 1, and the origin  absorbing.
On visiting a virgin site, the walker is pushed back to its previous position with 
probability $r$ (analogous to $p'<p$ in the generalized epidemic process).
One may think of this as being due to a reflector
that sits just beyond the maximal site that has yet been visited.
(Initially the reflector is at $x=2$.)
Asymptotic analysis of the master equation, and exact iteration of transition
probabilities for finite times, yield a variable survival exponent $\delta$.
One may distinguish two kinds of reflector.  A
{\it soft reflector} moves forward one step at every encounter with the walker,
even if the latter is reflected back.  
In  this case we find \cite{rwmpr}

\begin{equation}
\delta = \frac{1+r} {2}.
\end{equation}
A {\it hard reflector}, on the other hand, moves only if the walker
succeeds in occupying the virgin site, analogous to the spread of
activity in the generalized epidemic process.  Here the decay exponent
can be much larger:
\begin{equation}
\delta = \frac{1}{2(1-r)}.
\end{equation}
An interesting feature of the hard-reflector case is that 
for $r $ close to 1, decay of $P(t)$ looks
faster than power-law over many decades: the asymptotic power law
only becomes evident at very long times.
Variable exponents $\delta$ and $\eta$ are also found in the
case of compact DP with movable reflectors \cite{rddunp}.
If and how these findings apply to the more difficult question of
spreading in models with an infinite number of absorbing configurations,
and in the generalized epidemic process, remains to be seen.

\section{Summary}

Absorbing-state phase transitions are attracting increasing attention as
connections to SOC, spatiotemporal chaos, and interface dynamics are 
uncovered.  At the same time, various puzzles regarding universality and the
nature of scaling in models with an infinite number of absorbing
configurations remain unsolved, and can be expected stimulate
analysis and the search for improved theoretical and computational
methods.

\newpage
\noindent{\bf Acknowledgments}
\vspace{1em}

Much of the work described here grows out of ongoing collaborations with
Mikko Alava, Dani ben-Avraham, Miguel-Angel Mu\~noz, 
Alessandro Vespignani, and Stefano Zapperi.
I have also benefitted from discussions with M\'arcio Argollo de Menezes,
Hugues Chat\'e, Deepak Dhar, Peter Grassberger, Jos\'e Guilherme Moreira,
G\'eza \'Odor, and Carmen P. C. do Prado.
This work was supported by CNPq, Brazil.


\begin{thebibliography}{99}
\bibitem{rnote}
      I am not aware of a rigorous argument for the existence of a phase transition
      at a $\zeta_c$ strictly less than 1 in this class of model.
\end{thebibliography}
\end{document}